\documentclass[aps,prd,superscriptaddress,groupedaddress,nofootinbib,nobibnotes]{revtex4}

\usepackage{graphicx}
\usepackage{dcolumn}
\usepackage{bm}
\usepackage{amssymb}
\usepackage{epstopdf}
\usepackage{amsmath}
\usepackage{amsfonts}
\usepackage{color}
\usepackage{mathrsfs}

\newcommand{\be}{\begin{equation}}
\newcommand{\ee}{\end{equation}}
\newcommand{\ba}{\begin{eqnarray}}
\newcommand{\ea}{\end{eqnarray}}
\newcommand{\nn}{\nonumber}
\newcommand{\barr}{\begin{array}}
\newcommand{\earr}{\end{array}}

\newcommand\lsim{\mathrel{\rlap{\lower4pt\hbox{\hskip1pt$\sim$}}
        \raise1pt\hbox{$<$}}}
\newcommand\gsim{\mathrel{\rlap{\lower4pt\hbox{\hskip1pt$\sim$}}
        \raise1pt\hbox{$>$}}}

\def\Tr{\mbox{Tr}}

\begin{document}

\title{Fast Wiener filtering of CMB maps with Neural Networks}

\author{Moritz M\"unchmeyer}
\affiliation{Perimeter Institute for Theoretical Physics, Waterloo, ON N2L 2Y5, Canada}
\author{Kendrick~M.~Smith}
\affiliation{Perimeter Institute for Theoretical Physics, Waterloo, ON N2L 2Y5, Canada}

\date{\today}

\begin{abstract}
We show how a neural network can be trained to Wiener filter masked CMB maps to high accuracy. We propose an innovative neural network architecture, the WienerNet, which guarantees linearity in the data map. Our method does not require Wiener filtered training data, but rather learns Wiener filtering from tailored loss functions which are mathematically guaranteed to be minimized by the exact solution. Once trained, the neural network Wiener filter is extremely fast, about a factor of 1000 faster than the standard conjugate gradient method. Wiener filtering is the computational bottleneck in many optimal CMB analyses, including power spectrum estimation, lensing and non-Gaussianities, and our method could potentially be used to speed them up by orders of magnitude with minimal loss of optimality. The method should also be useful to analyze other statistical fields in cosmology.
\end{abstract}

\maketitle

\section{Introduction}
\label{sec:intro}

The cosmic microwave background (CMB) is the oldest radiation in the universe, propagating through the universe since recombination about 400.000 years after the big bang. At this moment in time the universe had cooled down sufficiently to become transparent to light. Because of this closeness to the big bang, the CMB is an ideal probe of the fundamental parameters of cosmology. For this reason, the CMB has been the driving force behind the era of precision cosmology, which allowed to determine fundamental cosmological parameters with percent level accuracy (see e.g the most recent results from the Planck experiment~\cite{Aghanim:2018eyx}). Reaching this level of precision requires sophisticated data analysis methods, and for each new generation of experiments improved methods have to be developed to fully exploit the data. There is however one basic building block of CMB data analysis that is usually required to make it statistically optimal: the process of Wiener filtering real and simulated CMB maps. This has become more and more difficult to achieve with larger and larger data sets, to the point that for the recent Planck satellite results many analyses have been performed without it, at the cost of sub-optimality.

Wiener filtering is the basic building block and computational bottleneck of the statistically optimal analysis of Gaussian random fields like CMB maps. Assuming the data vector $d$ is the linear sum of signal $s$ and noise $n$ with independent covariance matrices $S$ and $N$, the Wiener filtered data $d_{WF}$ is defined by
\be
d_{WF} = S (S+N)^{-1} d.  \label{eq:wf_def}
\ee
For a data set with $N$ pixels, the direct inversion of a dense $N \times N$ covariance matrix is impossible for current CMB maps with millions of pixels. Conjugate gradient solvers~\cite{Press:2007:NRE:1403886} are usually employed to perform Wiener filtering of CMB data (see e.g.~\cite{Smith:2007rg}) but the computational costs are enormous for Planck resolution and Wiener filtering a large ensemble of maps remains very difficult even with large computing resources.

In this paper we show how neural networks can be trained to Wiener filter CMB maps to a very good approximation, at a tiny fraction of the computational costs of the exact method. Our method does not require Wiener filtered maps as training data, but rather learns Wiener filtering from tailored loss functions which are mathematically guaranteed to be minimized by the exact solution. We employ an unusual neural network architecture, called the WienerNet, which guarantees that the output map is linear in the input map (but non-linear in the mask), unlike in conventional neural networks.

Our neural network solution to Wiener filtering can be used in the conventional pipelines of CMB data analysis. The massive speed up with respect to the standard method should allow us to perform previously intractable analyses. Our method is in particular interesting where large numbers of simulations have to be Wiener filtered (this is usually the case for an optimal analysis), or where Wiener filtering has to be performed many times in an iterative likelihood maximisation. On a more conceptual level, we demonstrate how neural networks can learn the most fundamental task of the analysis of Gaussian random fields. In this way we lay a foundation to employ neural networks reliably for more complicated tasks in cosmological data analysis. 

Neural networks have recently been used in cosmology in numerous other contexts. Specifically in CMB physics, reference~\cite{Caldeira:2018ojb} used neural networks to reconstruct the CMB lensing potential. The author's test setup of small CMB sky patches has inspired our own choices, although our learning task and neural network architecture are different. Neural networks have also been proposed for galaxy shear estimation~\cite{Ribli:2019wtw}, to generate galaxy distributions in a dark matter simulation~\cite{Zhang:2019ryt}, to learn cosmological structure formation~\cite{He:2018ggn}, for cosmological parameter estimation~\cite{Ravanbakhsh:2017bbi,Mathuriya:2018luj} and for likelihood-free inference~\cite{Alsing:2019xrx}. Contrary to some of these examples, here we learn a mathematically well defined task for which an optimal (although slow) solution is known, and can therefore precisely quantify the neural network's performance. 

The paper is organized as follows. In Sec. \ref{sec:lossfunctions} we describe the optimization task and appropriate loss functions. Then we discuss our neural network architecture in Sec. \ref{sec:neuralnetwork}. The input data and training process are described in Sec. \ref{sec:dataset}. Finally we present our results for temperature and polarisation Wiener filtering in Sec. \ref{sec:results}. We conclude in Sec. \ref{sec:conclusion} with a discussion of possible applications.

\section{Optimization problem and loss functions}
\label{sec:lossfunctions}

To train a neural network, a loss function must be specified.
In this section, we will study the question:
if we want the network to learn the Wiener filter in Eq.~(\ref{eq:wf_def}),
what loss function should we use?

\subsection{Definitions and notation}

Let $s$ denote the true CMB sky,
and let $d=s+n$ be the noisy data which is passed to the neural
network as input.
Let $y$ denote the output filtered map from the neural network. The neural network architecture is constrained to be linear in $d$
(for a fixed mask).  That is,
\be
y = M d  \label{eq:M_def}
\ee
for some matrix $M$ which is learned during training.
We would like the neural network to learn the Wiener filter:
\be
y_{\rm WF} = S (S+N)^{-1} d  \label{eq:ywf_def}
\ee
where $S,N$ denote signal and noise covariance matrices respectively.

In this paper, we assume that the noise covariance $N$ is diagonal
in pixel space, i.e.~the noise is assumed uncorrelated between pixels,
but is allowed to be anisotropic.
We represent the mask as a limiting case of anisotropic
noise, by taking the noise level to be infinity in masked pixels.
(In implementation, we set the corresponding entries of $N^{-1}$ to zero).

Note that we are using an index-free notation where CMB maps are
denoted with lower-case ($s,d,y,\cdots$), and operators are denoted
with upper-case ($S,N$).
This basis-free notation does not specify whether maps are represented
in pixel space or Fourier space, and in implementation we freely switch
between representations depending on which is more convenient.
For example, we simulate the CMB signal $s$ in Fourier space, but
simulate the noise $n$ in pixel space, since the corresponding covariance
matrices $S,N$ are diagonal in Fourier and pixel space respectively. The implementation will be described in more detail below.

\subsection{Loss functions for Wiener filtering}

During training, the weights of the neural network are varied
in order to minimize a prescribed loss function $J$.
The loss function is usually stochastic, i.e.~it is specified as a function 
$J(s,d,y)$, and the training process minimizes the expectation value
$\langle J(s,d,y) \rangle$, where the expectation value $\langle \cdot \rangle$
is taken over realizations $(s,d)$ in the training set.

We will say that a loss function $J(s,d,y)$ is ``viable'' if it satisfies the following property.
When $\langle J(s,d,y) \rangle$ is minimized over all choices of matrix $M$ in Eq.~(\ref{eq:M_def}),
then the minimum occurs for $M = S(S+N)^{-1}$.
Here, the expectation value $\langle \cdot \rangle$ is taken over an infinitely large training set.
Viability ensures that the Wiener filter can be learned to arbitrarily high accuracy,
provided that
the training set is large enough,
the number of training iterations is large enough,
and the network has enough capacity to represent the Wiener filter $M = S(S+N)^{-1}$.

There is more than one viable candidate for the loss function $J(s,d,y)$.
In this paper, we will consider three possibilities $J_1,J_2,J_3$.
We will give the definitions first, then show that each loss function is viable
in the sense just defined.

\begin{enumerate}

\item 
The loss function $J_1$ is defined by:
\be
J_1(d,y) = \frac{1}{2} (y - y_{\rm WF})^T A (y - y_{\rm WF})  \label{eq:J1_def}
\ee
where $A$ is an arbitrary positive definite matrix.  (We usually use $A=I$,
but we experimented with reweighting in $l$, by choosing $A$ to be diagonal
in Fourier space.)  Here, $y_{\rm WF} = S(S+N)^{-1} d$ is the Wiener filtered
map, which is a function of the input data realization $d$.

Minimizing $J_1$ simply corresponds to minimizing the mismatch between the neural
network output $y$ and the Wiener filtered map $y_{\rm WF}$.
This makes the training process intuitive, but computationally expensive since
we need to Wiener-filter each training realization using the conjugate gradient algorithm.
Indeed, avoiding the high computational cost of exact Wiener filtering is
the problem we are trying to solve in this paper!
For this reason, we do not advocate using $J_1$ in ``production'',
but we did find it useful during testing and debugging.

\item 
The loss function $J_2$ is defined by:
\be
\label{eq:j2}
J_2(s,y) = \frac{1}{2} (y-s)^T A (y-s)
\ee
where $A$ is an arbitrary positive definite matrix.
Training on $J_2$ corresponds to minimizing the mismatch between the neural
network $y$ and the true CMB sky $s$.

\item 
The loss function $J_3$ is:
\be
\label{eq:j3}
J_3(d,y) = \frac{1}{2} (y-d)^T N^{-1} (y-d) + \frac{1}{2} y^T S^{-1} y
\ee
This loss function is naturally motivated: we have
\be
J_3(d,y) = -\log P(s|d)_{s=y} + \mbox{const.}
\ee
where $P(s|d)$ is the posterior likelihood of true CMB sky $s$, given
noisy data realization $d$.
Thus, minimizing $J_3$ can be interpreted as training the network
to output the maximum likelihood CMB sky.
\end{enumerate}

Next we will show that each of these loss functions is viable,
in the sense that $\langle J \rangle$ is minimized for $M = S(S+N)^{-1}$.
For $J_1$, this is obvious from the definition~(\ref{eq:J1_def}),
since $J_1=0$ iff $M = S(S+N)^{-1}$ (so that $y = y_{\rm WF}$).

Considering $J_2$ next, we first compute $\langle J_2 \rangle$ as a function of $M$:
\ba
\langle J_2 \rangle 
  &=& \left\langle \frac{1}{2} (y-s)^T A (y-s) \right\rangle \nn \\
  &=& \left\langle \frac{1}{2} ((M-1)s)^T A ((M-1)s)
                 + \frac{1}{2} (Mn)^T A (Mn) \right\rangle \nn \\
  &=&  \frac{1}{2} \Tr\Big( (M-1)^T A (M-1) S \Big)
     + \frac{1}{2} \Tr\Big( M^T A M N \Big)
\ea
where we have used $y = M(s+n)$ in the second line.
Differentiating with respect to $M$, we get:
\ba
\frac{\partial \langle J_2 \rangle}{\partial M}
 &=& A (M-1) S + A M N  \nn \\
 &=& A M (S+N) - A S
\ea
Setting this to zero and solving for $M$, we get $M = S(S+N)^{-1}$,
completing the proof that $J_2$ is a viable loss function.

We do a similar calculation for $J_3$.
First we calculate $\langle J_3 \rangle$ as a function of $M$:
\ba
\langle J_3 \rangle 
  &=& \left\langle \frac{1}{2} (y-d)^T N^{-1} (y-d) + \frac{1}{2} y^T S^{-1} y \right\rangle \nn \\
  &=& \left\langle \frac{1}{2} ((M-1)d)^T N^{-1} ((M-1)d)
                 + \frac{1}{2} (Md)^T S^{-1} (Md) \right\rangle \nn \\
  &=&  \frac{1}{2} \Tr\Big( (M-1)^T N^{-1} (M-1) (S+N) \Big)
     + \frac{1}{2} \Tr\Big( M^T S^{-1} M (S+N) \Big)
\ea
Differentiating with respect to $M$:
\ba
\frac{\partial \langle J_3 \rangle}{\partial M}
 &=& N^{-1} (M-1) (S+N) + S^{-1} M (S+N) \nn \\
 &=& (S^{-1} + N^{-1}) M (S+N) - N^{-1} (S+N)
\ea
Setting this to zero and solving for $M$, we get:
\be
M = (S^{-1} + N^{-1})^{-1} N^{-1} = S (S + N)^{-1}
\ee
completing the proof that $J_3$ is viable.

Summarizing, all three loss functions $J_1,J_2,J_3$ are viable.
The $J_3$-loss seems most naturally motivated, but the $J_2$-loss has
the interesting property that it can be reweighted in $l$ (via the matrix $A$,
which is a free parameter), which could help convergence during training. 
Below we test the loss functions $J_2$ and $J_3$, and find the best results with $J_3$. 

\section{Neural network architecture}
\label{sec:neuralnetwork}

\begin{figure}
\centerline{\includegraphics[width=13cm]{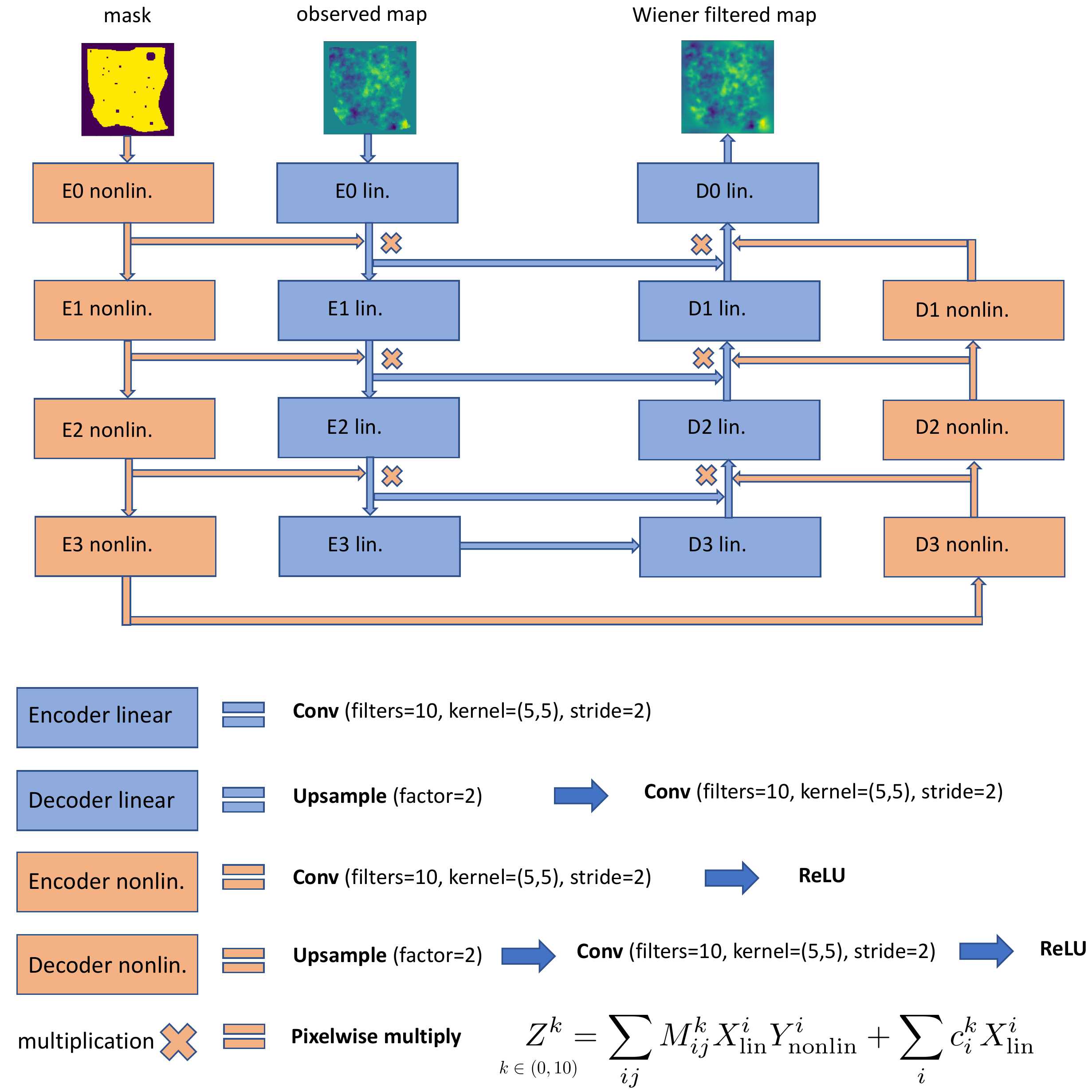}}
\caption{Top: Network architecture of the WienerNet with a non-linear (red) and linear (blue) path with encoders (E) and decoders (D). Skip connections in the non-linear path are not shown, but equivalent with those in the linear path. Our default architecture has two more layers of encoders/decoders than shown in this figure. Bottom: Definition of the building blocks. Encoders downsample by a factor of two (stride=2), and decoders upsample by a factor of two, except encoder 0 and decoder 0 which have stride 1.  The multiplication operation takes an input from the linear CMB path $X$ and the non-linear mask path $Y$ and multiplies them pixelwise according to the learned weights $M_{ij}^k$ and $c_i^k$.}
\label{fig:fig_nn_1and2}
\end{figure}

Neural networks parametrise a family of functions that must be well suited to learn the desired mapping from input to output data. Much of the impressive progress in deep learning (see e.g.~\cite{Goodfellow-et-al-2016}) in recent years has been to due to improved neural network architectures, which for example improve the behavior of gradients in deep networks (residual connections~\cite{2015arXiv151203385H}) or avoid overtraining (dropout layers~\cite{JMLR:v15:srivastava14a}). Our network architecture is inspired by techniques from deep learning image segmentation (where each pixel in an image is assigned a class value, e.g. corresponding to tree or car). Similar networks have been proposed for different tasks in cosmological data analysis, for example in~\cite{Zhang:2019ryt,Caldeira:2018ojb}. In this work however we crucially enforce linearity in the CMB data, which is very different from the highly non-linear functions in traditional neural networks. Our proposed neural network architecture, the WienerNet, is shown in Fig.~\ref{fig:fig_nn_1and2}. We now describe the main features of our architecture.

\medskip

\textbf{Convolutional Neural Network}. At the most basic level our neural network is a \textit{feed forward network}, in which information is propagated from the input through several functional building blocks called \textit{layers} to the output. The layers depend on a set of parameters called \textit{weights}, which are trained by gradient descent of the loss function, in a process called \textit{back propagation}. For images, i.e. data which is organized spatially, the main functional building block are convolutions, which provide spatially localized operations and vastly reduce the number of weights compared to more general functions. The input tensor of a convolutional layer has size $(X,Y,N)$, where $(X,Y)$ are to be understood as the image dimensions and $N$ is the number of \textit{feature maps}. The convolutional layer transforms the input tensor $L$ into an output tensor $L'$ of size $(X',Y',M)$ by applying a convolution as follows:
\ba
L'^{m}_{x'y'} = \sum_{ijm} L^n_{(x'-1)\times s+i, (y'-1)\times s+j} K^{nm}_{ij} + b^{m}
\ea
Here $K$ is the \textit{kernel} with kernel size $(I,J)$ and consists of  $N \times M \times I \times J$ weights, and $b$ is the bias with $M$ weights. The parameter $s$ is called the \textit{stride}, and for $s>1$ one obtains a \textit{downsampling convolution}, i.e. the output ``image'' has a lower resolution than the input image. All convolutions in this paper are of type \textit{valid}, i.e. we without zero padding. A convolutional neural network stacks many such convolutions on top of each other, usually with non-linearities in between them. For more details about convolutional neural networks see~\cite{Goodfellow-et-al-2016}.  For the task of Wiener filtering, convolutions are a natural choice since without a mask the Wiener filter is essentially a low-pass filter, which can be implemented in position space by a convolution.

\medskip

\textbf{Encoder-Decoder network of UNet type}. An important concept in machine learning are encoder-decoder architectures. The \textit{encoder} learns efficient representations of the input data, for example representing the semantic content of an input text, and the \textit{decoder} maps this representation to the desired output, for example a translation of the text in another language~\cite{2014arXiv1409.3215S}. Similarly, in image segmentation, the encoder learns a representation of the objects in the image, and the decoder maps it to a pixel segmentation (see e.g.~\cite{2015arXiv150504597R}). A successful architecture in this domain is the \textit{UNet}~\cite{2015arXiv150504597R}, which adds so called \textit{skip connections} which pass information directly from lower level encoders to lower level decoders. The UNet architecture is shown in the blue parts of Fig.~\ref{fig:fig_nn_1and2}. The encoders successively downsample the input map by a factor of two, using a convolution with stride two, while the decoders upsample the map by a factor of two and then apply a convolution with stride one.

For our Wiener filtering task, the heuristic motivation for the use of a UNet is that Wiener filtering CMB maps is a scale dependent process. To Wiener filter the large scales in the map, one does not need to know the small scales to good approximation. This fact was also exploited for the multigrid preconditioner in~\cite{Smith:2007rg}. In our implementation of the UNet, the encoders successively reduce the resolution by a factor of two, to get a more coarse grained view of the input map. The decoder path then learns how to assemble the information from different scales into an output map which contains all scales. To be able to analyze the largest scales in the map, we need to chain enough encoders so that at the highest level the convolutional layer sees the entire map in its receptive field.

\medskip

\textbf{Linear and non-linear path for data and mask}. The features described above are part of the standard deep learning image analysis toolkit developed over the last years. We now describe our innovation that makes them applicable to Wiener filtering. A key property of the Wiener Filter is that it is linear in the input map, which guarantees that a Gaussian field is transformed into a Gaussian field. Conventional neural networks have non-linearities called \textit{activation functions}, which build the highly non-linear functions needed in standard image analysis. In our architecture, we entirely drop activation functions in the map processing path (blue parts of Fig.~\ref{fig:fig_nn_1and2}), which is thus manifestly linear. The Wiener filter depends on the mask and noise of the experiment. In particular the mask breaks statistical homogeneity and thus translation invariance. To allow the network to learn this dependence on the mask, we process it in a second UNet type pathway (red parts of Fig.~\ref{fig:fig_nn_1and2}). The mask pathway can be non-linear, and we insert activation functions of \textit{ReLU} type (see e.g.~\cite{Goodfellow-et-al-2016}) after every convolution. The two pathways communicate from the non-linear to the linear layer (never the other way round), by the weight-dependent pixelwise multiplication defined in Fig.~\ref{fig:fig_nn_1and2}. In our setup we use a fixed mask for all examples in training and evaluation, so that the mask path does not perform a dynamic calculation at evaluation time, but rather provides an efficient functional representation of the influence of the mask on the WienerFilter. In a generalisation of our setup one could train with randomized masks to potentially learn a WienerFilter as a function of the mask. Our current setup however covers the important practical case of a fixed mask for a given experiment.

\medskip

A more detailed description of the WienerNet is provided in Appendix~\ref{app:nnarch}. We have also experimented with some basic modifications of our architecture. We confirmed that the non-linear mask path indeed substantially improves the results with respect to a linear mask path or no mask path at all. We also confirmed that the number of feature maps in the convolutions should be larger than one or two, and have chosen 10 for temperature and 16 and 32 for polarisation.

\section{Data set and training}
\label{sec:dataset}
We are now prepared to discuss the technical details of our neural network Wiener filtering pipeline, including data sets, neural network implementation and training process.

\subsection{CMB data and exact Wiener filtering}

Our training data consists of simulated CMB temperature and polarization maps. The CMB temperature field $T$ is a real valued scalar field describing the temperature in each pixel. CMB polarisation is usually expressed in terms of the real valued Stokes parameters $(Q,U)$, which describe the polarisation state of the CMB light in each pixel. Technically, the complex linear polarisation $P_\pm$ is given by $P_\pm = Q \pm i U$ and is said to be \textit{spin-2} due to its rotation properties. Alternative to the $(Q,U)$ basis, one can construct a basis from scalar fields called $(E,B)$, which are loosely related to the more familiar divergence and rotation for vector fields in Euclidean space. This basis has the advantage that it is the natural basis for physical predictions of the associated signal power spectra. Note that the transformation of $(Q,U)$ to $(E,B)$ is non-local, and we only use the $(E,B)$ basis in Fourier space. We refer the reader for example to references~\cite{Hu:1997hv,Lewis:2006fu} for a detailed discussion of CMB polarisation. 

The training, validation and test data set for both temperature and polarisation were generated with the {\tt quicklens}\footnote{https://github.com/dhanson/quicklens} package. In our default configuration, we simulate $(T,Q,U)$ sky patches of size 5 deg $\times$ 5 deg, at a resolution of 128 $\times$ 128 pixels. This is small enough so that we can use the flat sky approximation, where spherical harmonics can be replaced with ordinary Fourier transforms. The CMB maps are generated with current best fit cosmological parameters (with tensor-to-scalar ratio $r=0$) and include lensing. We have added different realistic experimental noise levels, specified below. We have generated a somewhat arbitrary mask, without any particular symmetry, representing sky cuts and smaller point-like source, which can be seen in Fig.~\ref{fig:maps_comparison_t} (right top). In total we create 10000 training maps as well as 1000 validation maps for each training setup. One could easily create much more training data, but we found good convergence with this number of maps. An interesting option would be to create maps directly during the training process and to use each map only once, which would eliminate the possibility of overtraining (to be discussed below). Map making in our setup is computationally relatively cheap, but with our implementation this ``online learning'' slowed down the training process by a factor of a few, so we opted for a static training set.

To test the quality of our neural network Wiener filter (not for its training), we compare it to the exact answer obtained from an iterative conjugate gradient method. This method is also implemented in {\tt quicklens}, and uses the multigrid preconditioner proposed in~\cite{Smith:2007rg}. We generate an independent test data set of 300 Wiener filtered maps for each setup.

\subsection{Implementation of the loss functions}

In the case of temperature Wiener filtering, the input data consist of the pair $(T^{\mathrm{obs}},M)$, where $M$ denotes the mask (a map of with elements one and zero) and is the same for each example in the training data in our setup. To evaluate the loss function $J_2$, we need the true map $T^{\mathrm{sky}}$, which could be considered the label of our training data set. The $J_2$ loss can be evaluated either in pixel space or in Fourier space, and we have experimented with both cases. We have found that the simplest possible case works well, i.e. evaluating the loss in pixel space and setting $A=\mathcal{I}$ in Eq.~\ref{eq:j2}. In this case the loss function is the simple sum over squared differences of pixels, i.e.
\ba
J_2^{T} = \sum_i^{n_{\mathrm{pix}}} (T^{\mathrm{NN}}_i - T^{\mathrm{sky}}_i)^2
\ea
To evaluate the loss function $J_3$, the required CMB map is the input map $T^{\mathrm{obs}}$ itself, i.e. we are not performing ordinary supervised learning with labels. For this loss, some theoretical input is needed, the signal and noise power spectra of the observed CMB data. This is no restriction as the same quantities are needed to generate the training data. The first term is naturally evaluated in pixel space, while the second term is naturally evaluated in Fourier space:
\ba
J_3^{T} = \sum_i^{n_{\mathrm{pix}}} \frac{(T^{\mathrm{NN}}_i - T^{\mathrm{obs}}_i)^2}{N_i} + \sum_{\bm{\ell}} \frac{ T^{\mathrm{NN}}_{\bm{\ell}} T^{\mathrm{NN}*}_{\bm{\ell}} }{C^T_\ell}
\ea
In the first term, $N_i$ is the pixel noise variance of the experiment within the observed sky, and infinity in the masked parts. The second term sums over all Fourier modes (up to the Nyquist frequency associated with the pixelisation) and weights them by the signal power spectrum $C^T_\ell$ of the CMB sky. Finally, we have also used the combined loss function
\ba
J_4^{T} = J_2^{T} + a J_3^{T}.
\ea
with $a$ a constant adjusted empirically to make both terms roughly equally large.

In analogy with the temperature case, for polarisation we have 
\ba
J_2^{QU} = \sum_i^{n_{\mathrm{pix}}}  (Q^{\mathrm{NN}}_i - Q^{\mathrm{sky}}_i)^2  + (U^{\mathrm{NN}}_i - U^{\mathrm{sky}}_i)^2
\ea
and 
\ba
J_3^{QU} = \sum_i^{n_{\mathrm{pix}}} \frac{(Q^{\mathrm{NN}}_i - Q^{\mathrm{obs}}_i)^2}{N_i} + \frac{(U^{\mathrm{NN}}_i - U^{\mathrm{obs}}_i)^2}{N_i}  + \sum_{\bm{\ell}} \frac{ E^{\mathrm{NN}}_{\bm{\ell}} E^{\mathrm{NN}*}_{\bm{\ell}} }{C^E_\ell} +  \frac{ B^{\mathrm{NN}}_{\bm{\ell}} B^{\mathrm{NN}*}_{\bm{\ell}} }{C^B_\ell}
\ea
The second loss term here is evaluated in the $E,B$ basis described above, since this is the basis in which physical predictions for the signal power spectra are conveniently expressed.

\subsection{Neural network implementation and training}

We have implemented our neural network in TensorFlow~\cite{tensorflow2015-whitepaper}, which automatically provides an efficient GPU implementation. TensorFlow also provides FFTs in its computation graphs, which is necessary to implement our loss functions. Our default architecture uses six encoder/decoder pairs (two more than shown in Fig.\ref{fig:fig_nn_1and2}), each of which has kernel size 5 $\times$ 5 as well as 10 feature maps. In total, our network has about $80.000$ parameters, which makes it small compared to most modern deep neural networks with millions of parameters.  

As we will be working partially in Fourier space, we need periodic boundary conditions. This allows a fair comparison with {\tt quicklens} Wiener filtered maps, which also impose periodic boundary conditions. To analyse real data, one would either use an implementation on the sphere for full sky observations, or, for partial sky observations, make the masked area around the measured area large enough so that the Wiener filtered map falls to zero on all sides. This is independent of whether one uses a neural network or the exact method. It would be ideal to implement the periodic boundary conditions directly on the level of the convolutions in TensorFlow. Since this would potentially require deep changes in the code, we have opted for a simpler method which achieves equivalent results, but at the cost of more computation: We pad both input map and mask periodically, i.e. we enlarge the input data in all directions in a periodic way. For the highest layer convolutions to be able to see the entire map, one roughly needs to feed an input map with twice the side length of the original map. The stack of valid-mode convolutions in the network successively reduces the enlarged input map to the correct output size. We provide the exact side lengths of all layers in appendix~\ref{app:nnarch}.

To train the network we use the popular Adam optimizer~\cite{2014arXiv1412.6980K}, using mini batches of size eight. We have found that an Adam learning rate of $10^{-4}$ gives good convergence ($10^{-3}$ is too large and $10^{-5}$ is too small). The network trains out in about 1000 epochs, where one epoch is defined as having used all $10.000$ training maps once for a gradient update. Training takes of order 30 hours on our GeForce GTX 1080 Ti graphics card (however good results are already achieved much earlier in the training process). We will show examples of the training and validation loss curve below. Analyzing the 300 test maps with the trained out neural network takes a few seconds, while exactly Wiener filtering them takes a few hours. This is a massive speed advantage of a factor of 1000 or more and the main advantage of the neural network.

\section{Results}
\label{sec:results}
In this section we first present the results for Wiener filtering temperature maps, for two different sizes of sky patches. We then discuss the results for polarisation, and compare different loss functions.

\subsection{Wiener filtering temperature maps}

\begin{figure}
\centerline{\includegraphics[width=12cm]{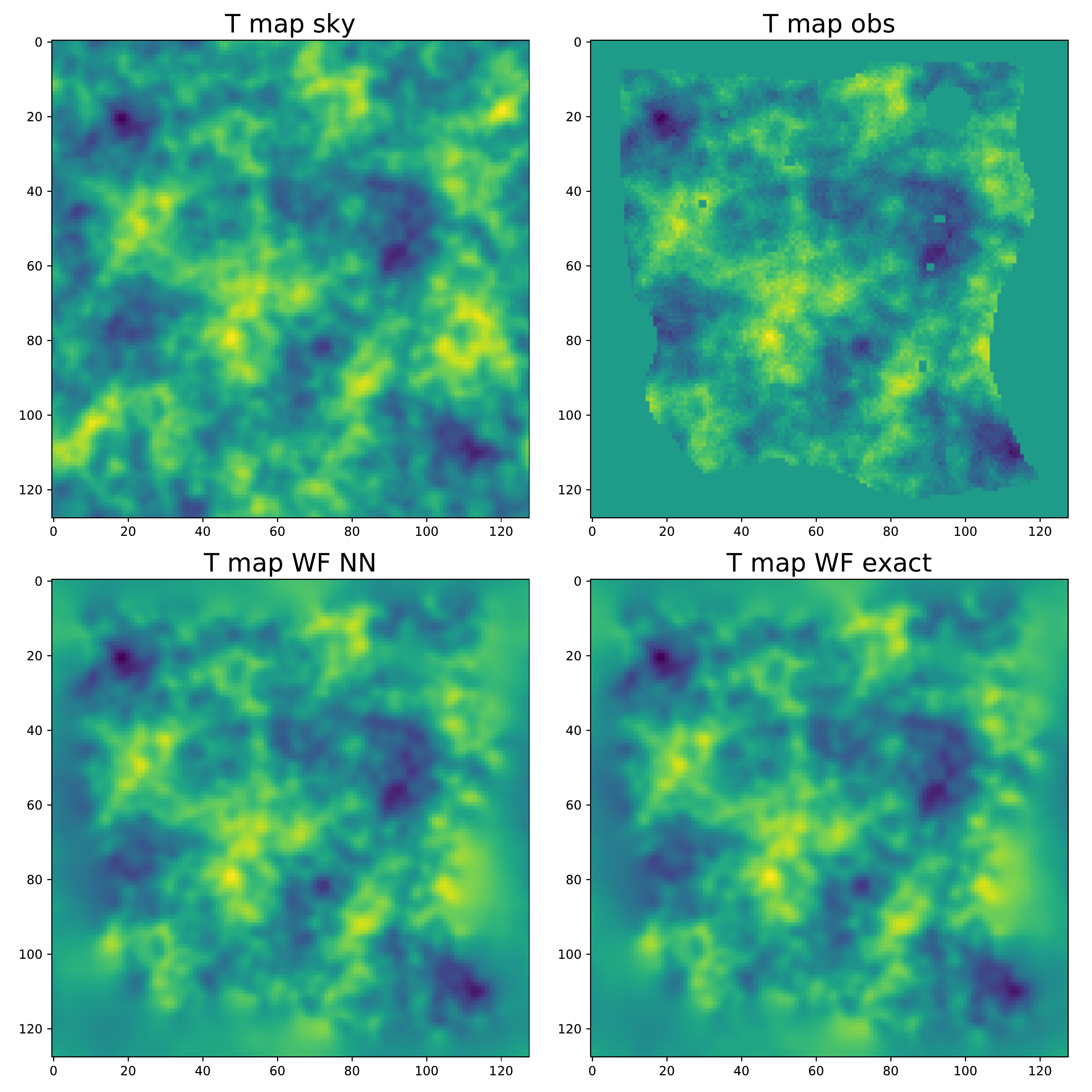}}
\caption{Randomly chosen example maps from the CMB temperature test set. We show sky map, observed map, neural network map and exact Wiener filtered map. The sky patch is of size $5\times 5~ \mathrm{deg}^2$ and pixelized with $128\times 128$ pixels. The neural network map and the exact Wiener filtered map show excellent visual agreement.}
\label{fig:maps_comparison_t}
\end{figure}

We start our discussion of the results by visually inspecting a randomly chosen test map in Fig.~\ref{fig:maps_comparison_t}. On the top left, the true sky map is shown, i.e. the real CMB fluctuations that an optimal noiseless mask free experiment would measure. On the top right, we show the experimental map, which was masked and has clearly visible Gaussian noise ($35\mu$K-arcmin) in it. Next, on the bottom left, we show the Wiener filtered map as it is obtained by our our proposed neural network after training with the $J_3$ loss function. On the bottom right, the exact Wiener filtered map from the Quicklens conjugate gradient computation is shown. Comparing the neural network with the exact answer, the visual correspondence is striking. 

\begin{figure}
\centerline{\includegraphics[width=12cm]{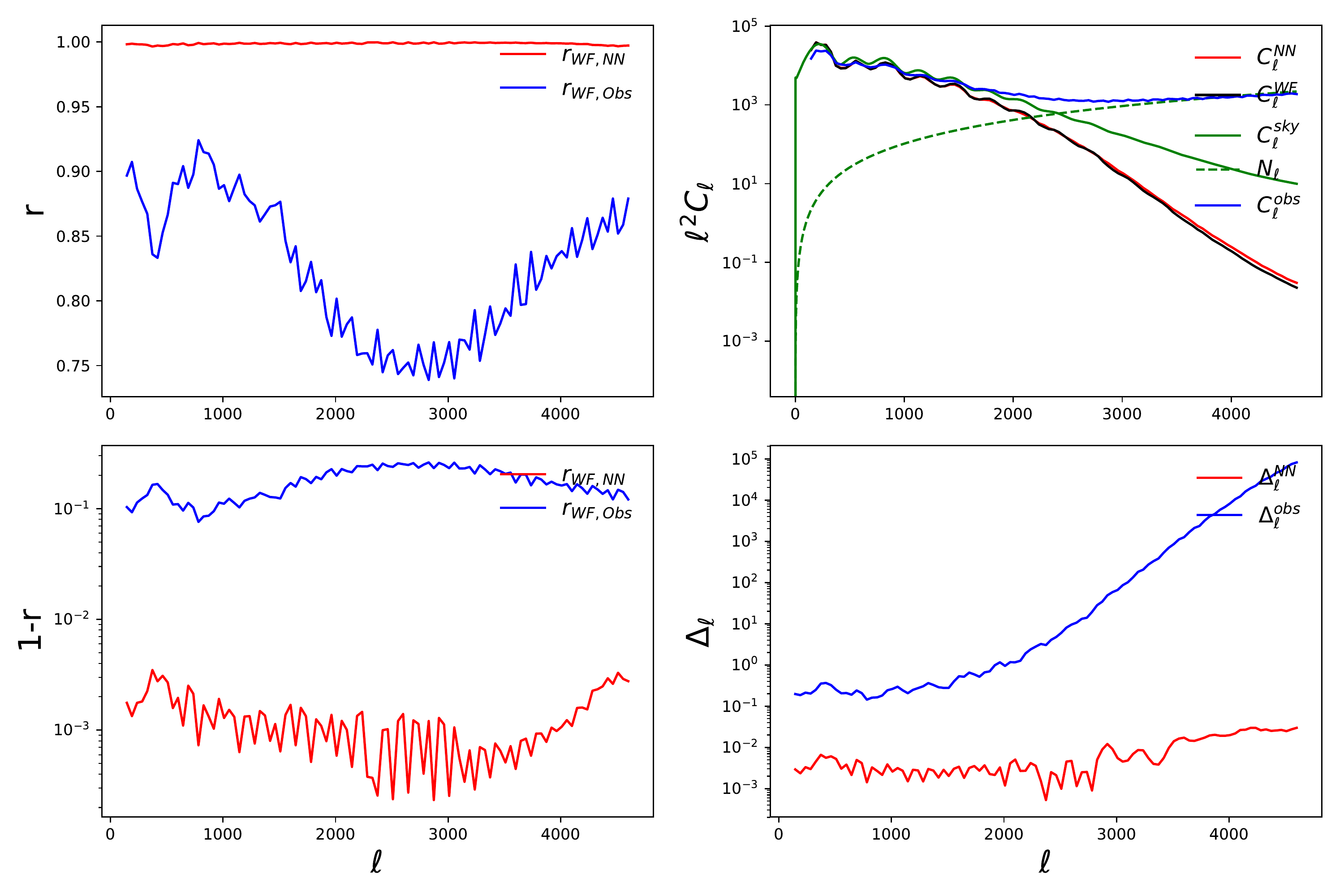}}
\caption{Quality measures for filtering 128x128 pixel CMB temperature maps with $35\mu$K-arcmin noise, obtained from 300 test maps. The cross-correlation coefficient between the neural network output and the exact Wiener filter is larger than $99\%$ on all scales.}
\label{fig:quality_measures_t}
\end{figure}

To establish the quality of our Wiener filtering, we show two measures, the cross correlation coefficient and the power spectrum, in Fig.~\ref{fig:quality_measures_t}. The cross correlation coefficient $r_\ell$ as a function of multipole $\ell$ is defined as
\be
r^{NN,WF}_\ell = \frac{ \left< a_{NN}(\bm{\ell}) \, a^*_{WF}(\bm{\ell}) \right> }{ \left(  \left< a_{NN}(\bm{\ell}) \, a^*_{NN}(\bm{\ell}) \right> \left< a_{WF}(\bm{\ell}) \, a^*_{WF}(\bm{\ell}) \right> \right)^{1/2}}
\ee
where $a_{NN}(\bm{\ell})$ and $a_{WF}(\bm{\ell})$ are the discrete fourier coefficients of the neural network output and the exact Wiener filtered map respectively. Note that the correlation coefficient cannot be improved by a simple $\ell$ dependent reweighting, and that for an unmasked observed map it is unity by definition. We also show the power spectra $C_\ell$ of the maps as well as the power of the difference map normalized with respect to the true power, i.e.
\be
\Delta_\ell = \frac{C^{NN-WF}_\ell}{C^{WF}_\ell}
\ee

The cross correlation coefficient between the neural network result and the exact Wiener filtering is shown in Fig.~\ref{fig:quality_measures_t} (top left and bottom left), extracted from averaging over 300 test maps. We show the entire $\ell$ range of the sky patch from the lowest Fourier mode up to the Nyquist frequency of the sky patch ($\ell_{Nq}=4608$ for the $5^\circ$ side length). The quality of Wiener filtering is excellent, with at least $99 \%$ cross correlation on all scales, independent of whether they are signal or noise dominated. It is likely that a larger neural network and a longer training would improve the results even further, but the degree of Wiener filtering reached here is sufficient for a near optimal cosmological analysis. For comparison, the plot also shows how well the observed (unfiltered) map is cross correlated with the exact Wiener filtered map. 

The power spectra of the maps are shown in Fig.~\ref{fig:quality_measures_t} top right. We find excellent agreement between the neural network and the exact Wiener filter, with a mild disagreement only appearing above $\ell \simeq 3500$, where the signal is already a factor of $1000$ lower than the noise. We show the power in the difference map of neural network and exact Wiener filter in Fig.~\ref{fig:quality_measures_t} bottom right. Up to $\ell \simeq 3500$ the power in the difference map is at the $1\%$ level or below.

\begin{figure}
\centerline{\includegraphics[width=12cm]{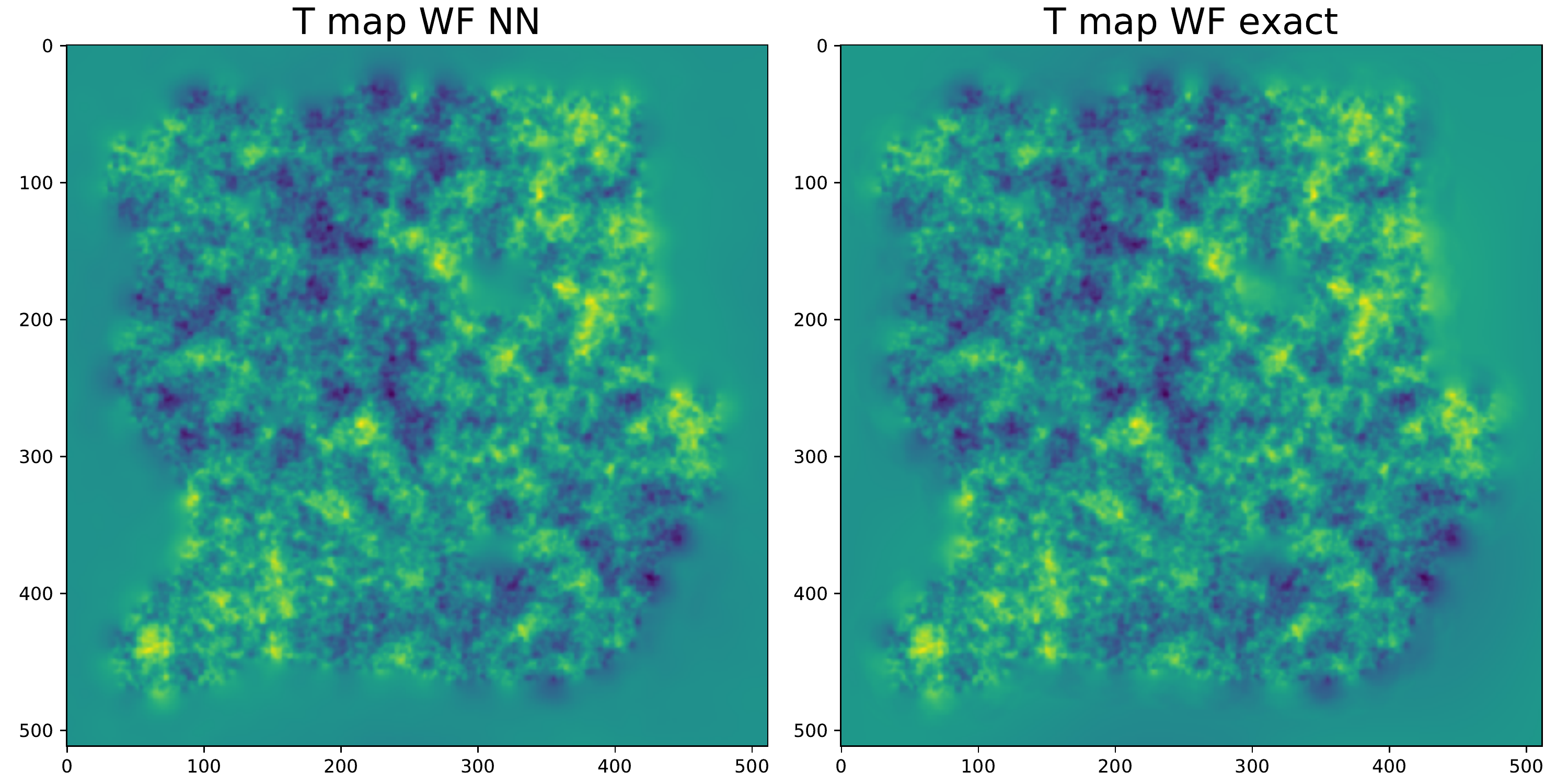}}
\centerline{\includegraphics[width=12cm]{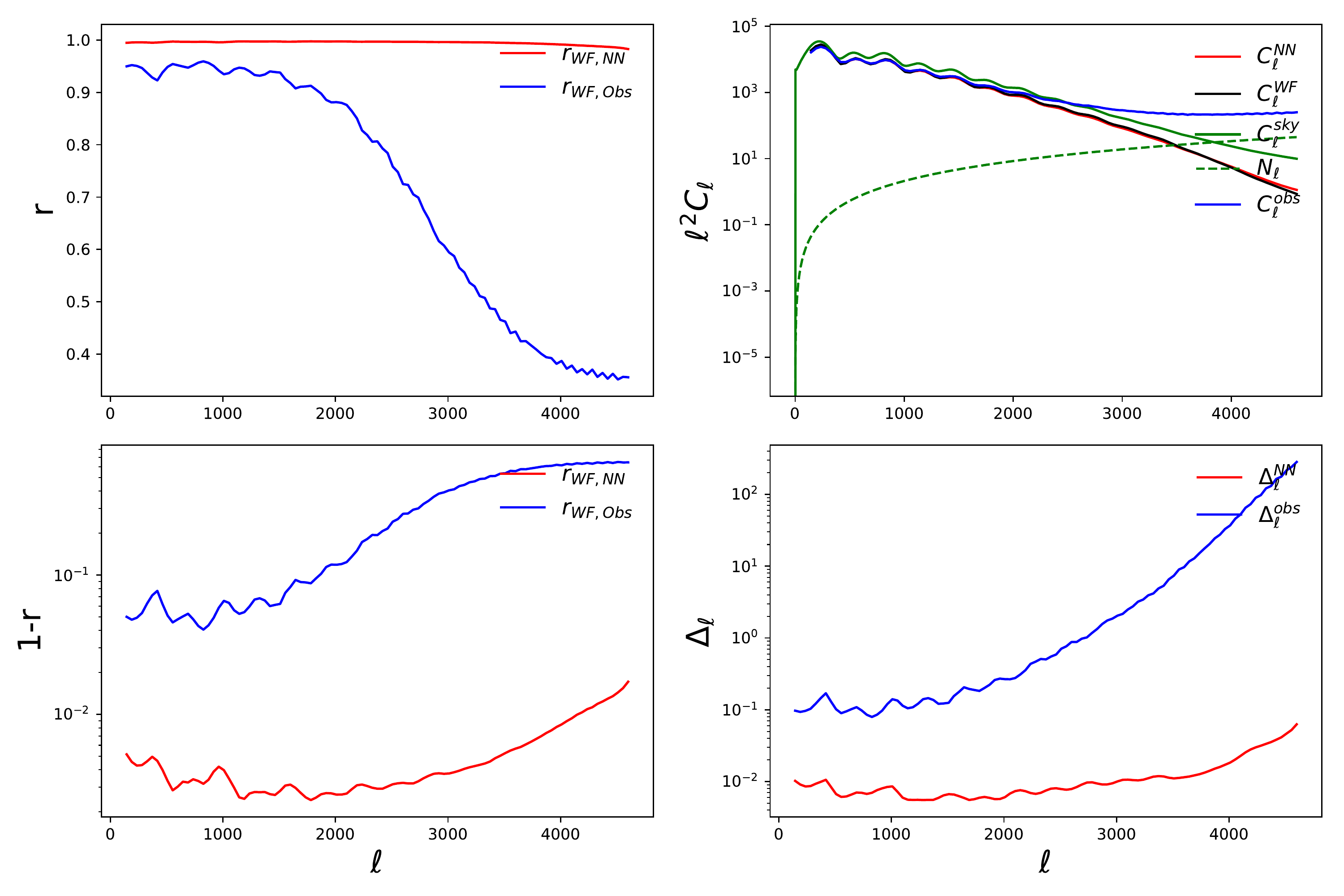}}
\caption{Upscaling the temperature Wiener filtering problem from 128x128 pixels to 512x512 pixels, using a different mask and $5\mu$K-arcmin noise. Again the visual and quantitative agreement between the neural network and the exact Wiener filter is excellent.}
\label{fig:quality_measures_t_large}
\end{figure}

Modern CMB experiments have millions of pixels. While we found the small patch of 128x128 pixels convenient for method development, we do not expect any fundamental computational or technical obstacle with treating realistic sizes. As a test of scalability, we have also trained a WienerNet with size 512x512, covering 20 deg $\times$ 20 deg with a different mask and noise ($5\mu$K-arcmin). The results are shown in Fig.~\ref{fig:quality_measures_t_large} and confirm the results for the smaller patch. In terms of network architecture, doubling the side length of the map requires to add one more pair of encoder/decoders to the WienerNet. 

\subsection{Wiener filtering polarisation maps}

\begin{figure}
\centerline{\includegraphics[width=12cm]{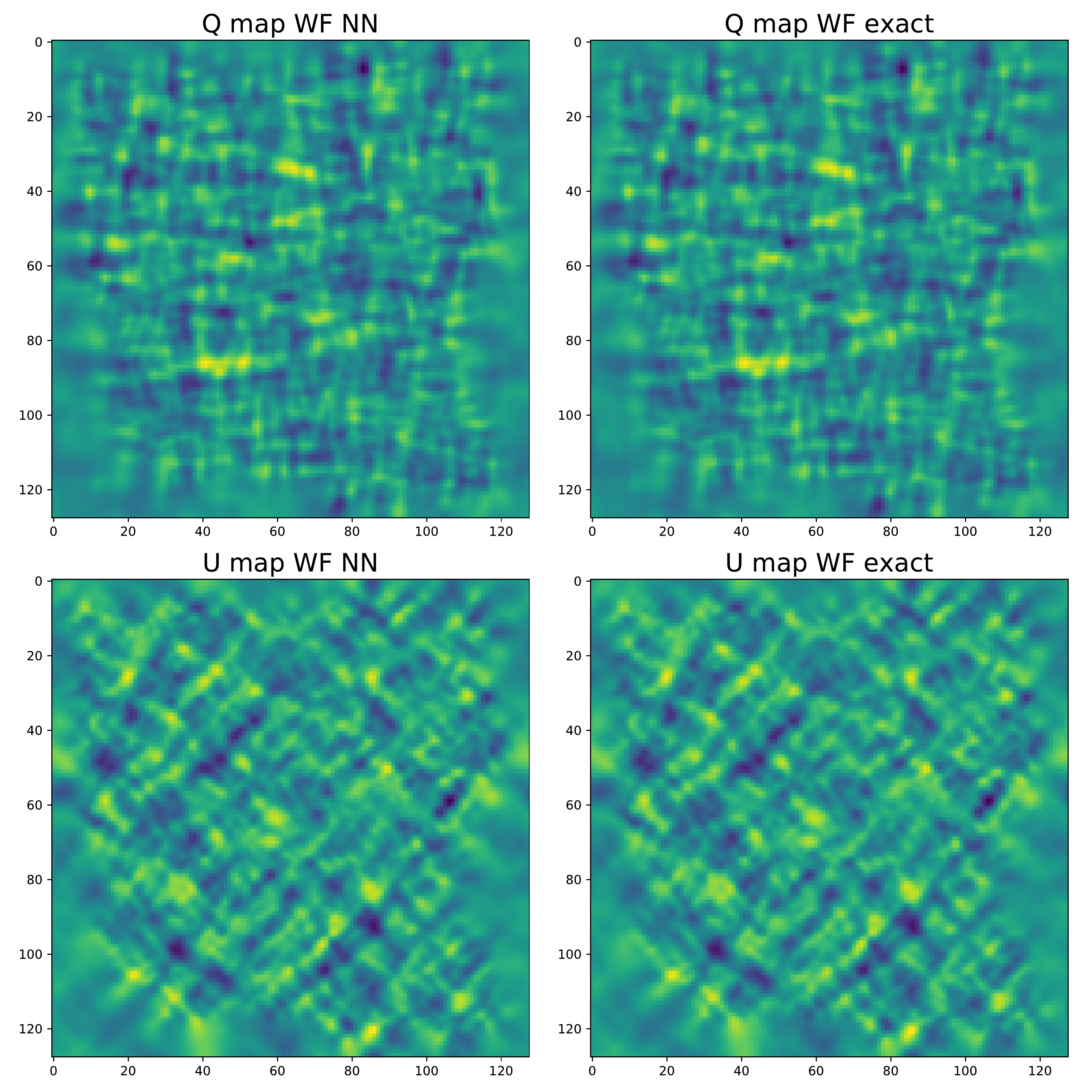}}
\caption{Randomly chosen example maps from the CMB polarisation test set.  We compare neural network output maps $(Q,U)$ and corresponding exact Wiener filtered map. As in the temperature case, the neural network output closely resembles the exact result.}
\label{fig:maps_comparison_qu}
\end{figure}

Encouraged by the excellent results in the temperature case we examine the Wiener filtering of polarisation maps $(Q,U)$. This is potentially challenging as it involves to learn the transformation $(Q,U)$ to $(E,B)$ in the presence of a mask. Further, the E-modes and B-modes have a very different signal-to-noise, as B-modes only arise from the lensing of primordial E-modes (in this analysis the tensor-to-scalar ratio is $r=0$, in line with current constraints). We show a randomly chosen test map in Fig.~\ref{fig:maps_comparison_qu} in the $Q,U$-basis. We have used the same mask as in the 128x128 pixel temperature example. The neural network results again resemble the exact Wiener filtering very closely.

\begin{figure}
\centerline{\includegraphics[width=12cm]{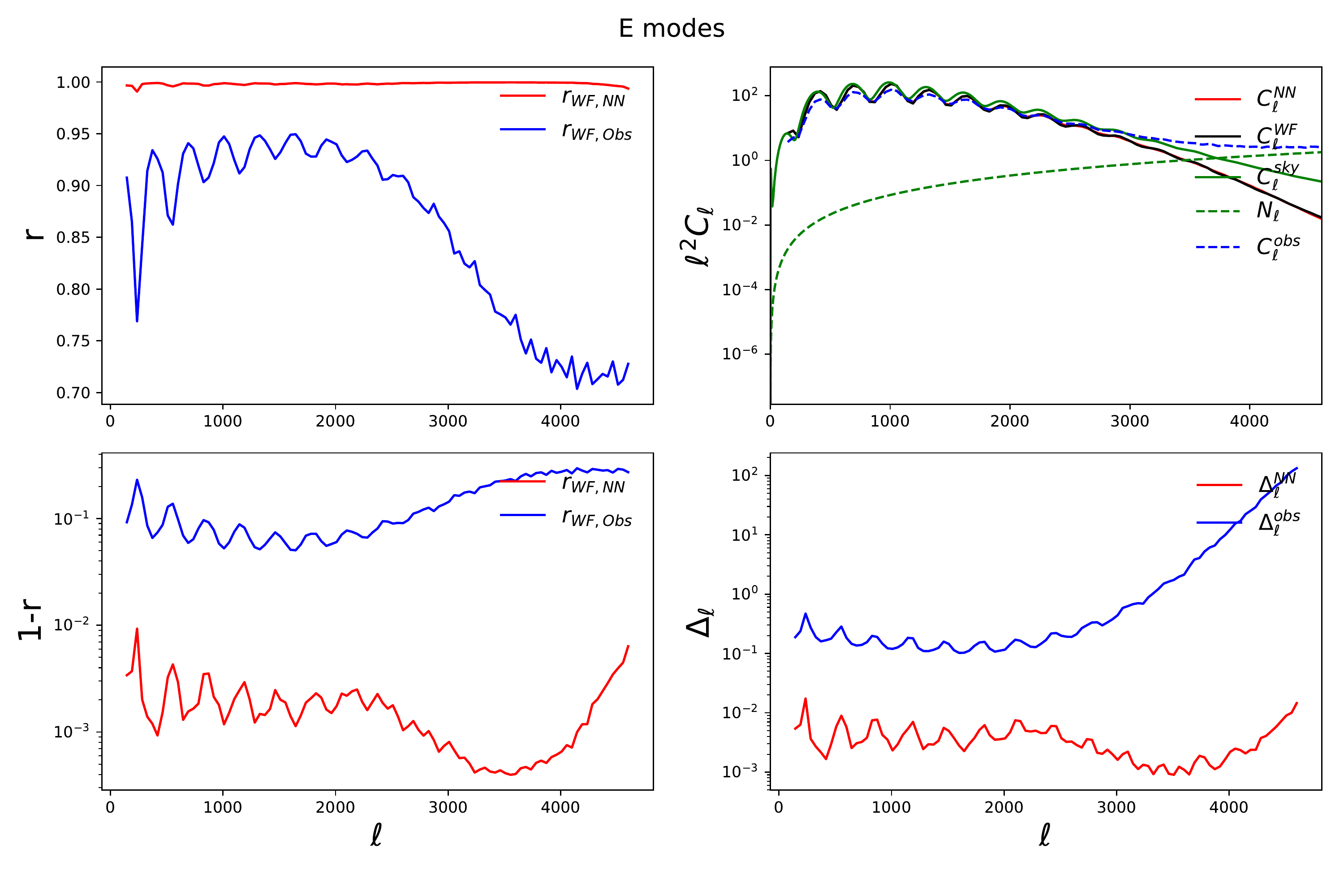}}
\centerline{\includegraphics[width=12cm]{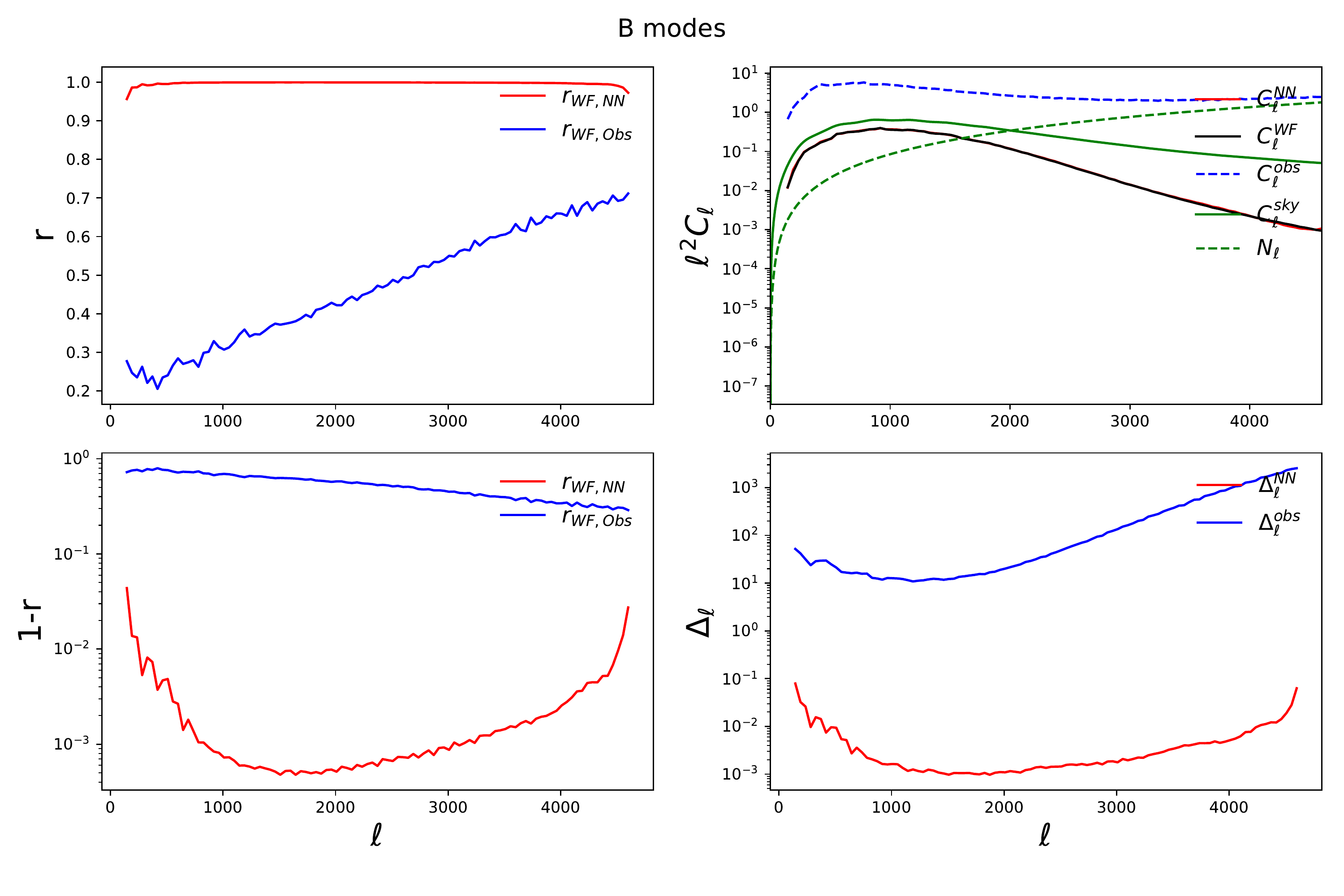}}
\caption{Quality measures for E-modes (top) and B-modes (bottom) for Wiener filtering 128x128 pixel CMB polarisation maps with $1\mu$K-arcmin noise, obtained from 300 test maps. For E-modes the correlation is $99\%$ or better on all scales. For B-modes the correlation is above $99\%$ or better in most of the $\ell$ range, with a drop to $95\%$ at the lowest $\ell$.}
\label{fig:quality_measures_eb}
\end{figure}

The natural basis to evaluate the quality of our Wiener filtering is $(E,B)$, since this is the basis that is used for physical analysis, e.g. cosmological parameter estimation. We plot our quality measures in Fig.~\ref{fig:quality_measures_eb}. We have obtained more than $99\%$ cross correlation with the exact Wiener filtered map over the entire $\ell$ range in E, and over most of the $\ell$ range in B with the exception of very low $\ell$. Small oscillations at the $<1\%$ level in the E map cross-correlation are induced by the baryon acoustic oscillations in the E-mode signal and are therefore absent in the B-modes. These small oscillations will not translate in a $1\%$ error in the power spectrum estimation but only in a small suboptimality of the error bar, as one would determine $\mathcal{L}(\hat{C}_\ell^{\mathrm WF} | C_\ell^\mathrm{true})$ from simulations. Fig.~\ref{fig:quality_measures_eb} also shows the power spectra and difference power spectra for E-modes and B-modes. For most of the $\ell$-range the difference power spectrum is again in the range of $1\%$ of the total power. 

For B-modes at very low and very high $\ell$ the correlation drops to about $95\%$. At high $\ell$ we are close to the Nyquist frequency $\ell_{Nq} = 4606$ of the patch, and it seems very likely that a higher resolution pixelisation would improve the behavior. Also the signal to noise in this region is very small. On the low $\ell$ side we note that the small sky patch of 5 deg $\times$ 5 deg has only a few Fourier modes in this range. It is possible that one could find further improvements of network parameters and training to reach $99\%$, for example by giving the non-linear path more capacity.

\subsection{Performance of different loss functions}

\begin{figure}
\centerline{\includegraphics[width=12cm]{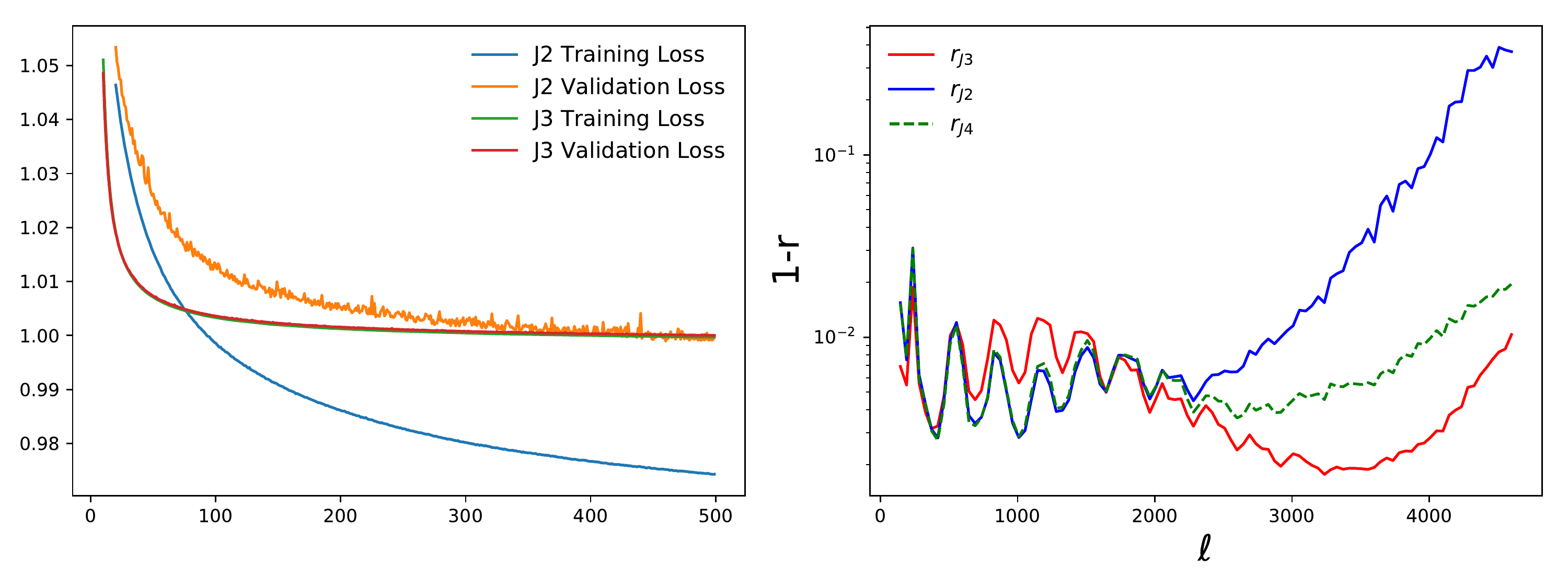}}
\caption{Comparison of the different loss functions for the polarisation case for epoch 20 to 500. Left: $J_2$ and $J_3$ training and validation loss, normalized so that the final validation loss is unity for both losses. For $J_3$ the two curves lie almost on top of each other, while $J_2$ shows some overtraining. Right: Best fit correlation coefficient for E-modes after 500 epochs. Here we also show $J_4$ which is defined as the sum of the two losses. In summary $J_3$ is the preferred loss function.}
\label{fig:loss}
\end{figure}

We have several mathematically valid loss functions, and it is interesting to compare their performance. The training and validation loss are shown in Fig.~\ref{fig:loss} (left) for $J_2$ and $J_3$, for the same polarisation setup as in the previous section. We have normalized the curves so that the validation loss in both cases is unity at the end of the training. One striking difference is that the $J_2$ loss function seems much more prone to overtraining, where training and validation loss diverge. This is likely because the $J_2$ loss function knows the true underlying CMB signal $s$, which contains information that cannot be obtained from the data $d$, while $J_3$ only has access to the noisy data realisation $d$. Further, it appears that the $J_2$ loss function has larger fluctuations (both cases use the Adam optimizer with learning rate $10^{-4}$). 

Decisive for the choice of the loss function is the performance on the test data. This is shown in Fig.~\ref{fig:loss} (right), at the example of the E-mode correlation coefficient (other quality measures behave similarly). A second striking difference between $J_2$ and $J_3$ is that the $J_3$ loss function performs much better at high noise. This is likely because in the $J_3$ case the loss function uses the noisy observed map in the square difference term. On the other hand $J_2$ seems to perform slightly better at large scales. We have therefore tested the combined $J_4 = J_2 + a J_3$, with $a$ set empirically so that the two losses are of the same order when the network trains out. As the plot shows this can be a valid alternative to $J_3$, but is inferior at high $\ell$. 

For these reasons, we have used the $J_3$ noise function for the results in the previous sections. Note that all three loss functions here are not entirely trained out and we have trained further epochs with the $J_3$ loss function for our final results presented in the previous section. Further, in this section we have used a ten filter WienerNet (as in the temperature case) to reduce training time.

\section{Conclusions}
\label{sec:conclusion}
In this paper we have demonstrated that convolutional neural networks can be trained to perform approximate Wiener filtering on masked CMB maps with high accuracy. Our main insights are the WienerNet architecture, which guarantees linearity in the CMB data, as well as finding and testing physical loss functions which are minimized by the desired solution, rather than using ordinary supervised learning. After training, the neural network Wiener filters a map by about a factor of 1000 faster than the exact conjugate gradient method (depending on choices like the network architecture and the convergence criterion of the exact method). This massive speed improvement could allow us to make key CMB data analysis much more efficient, and solve problems that were previously thought computationally intractable. 

An interesting generalisation would be to train the neural network on a set of randomized masks. Since the mask is an input of the neural network, it can in principle learn mask dependent Wiener filtering. This would be useful for experiments that use different masks for different analyses. Another obvious generalisation is to include inhomogeneous noise, which could be easily added to the mask path. It might also be possible to use the neural network Wiener filter result to precondition an exact conjugate gradient solver, and thereby obtain the exact Wiener filter with much less iterations than with ordinary preconditioners. 

A direct application of our ultra fast Wiener filtering would be CMB power spectrum estimation. A near optimal method to do so is to estimate the $\hat{C}_\ell^{\mathrm WF}$ from the Wiener filtered data map, and estimate the likelihood $\mathcal{L}(\hat{C}_\ell^{\mathrm WF} | C_\ell^\mathrm{true})$ from Monte Carlo. This approach has been successfully employed by the WMAP team in their CMB power spectrum analysis~\cite{2013ApJS..208...20B}. Estimating this likelihood was a computationally extremely challenging task for WMAP, and would be even harder for more high resolution data. Our fast Wiener filtering seems to be optimally suited for this task. A related and more challenging direction would be to estimate the cross power spectrum between different maps with different masks. Wiener filtering is also required for non-Gaussianity estimation. Here again it is necessary to Wiener filter a large set of Monte Carlo maps to estimate biases induced by experimental properties. The WienerNet is guaranteed not to induce artificial non-Gaussianity due to its linearity in the CMB data.

A very interesting possible application could be for CMB lensing power spectrum and de-lensing analysis. In principle optimal likelihood methods have been developed in~\cite{Carron:2017mqf,Millea:2017fyd} but they are again computationally very challenging, with potentially slow convergence, and have not been demonstrated on real data. Iteratively repeated Wiener filtering is a central building block in these methods, and we are currently examining whether the WienerNet is useful in this context.

The main missing component of our pipeline for an application to real data on a large sky fraction is an implementation on the sphere. Spherical convolutional neural networks have been recently implemented in~\cite{2018arXiv180110130C,2018arXiv181012186P}, which could perhaps be used to implement the WienerNet. Finally, the analysis of statistical fields appears in many other areas of cosmology besides the CMB, and the WienerNet technique will likely be useful in these areas too.

\section*{Acknowledgements}

Research at Perimeter Institute is supported by the Government of Canada
through Industry Canada and by the Province of Ontario through the Ministry of Research \& Innovation.
KMS was supported by an NSERC Discovery Grant and an Ontario Early Researcher Award. This research was enabled in part by support provided by Compute Ontario (www.computeontario.ca) and Compute Canada (www.computecanada.ca).”

\bibliography{wienerfiltering}

\appendix
\section{Detailed neural network architecture}
\label{app:nnarch}

In this appendix we provide the details necessary to reproduce our neural network setup. The network architecture and building blocks are defined in Fig.\ref{fig:fig_nn_1and2}. All encoders and decoders use a single convolution, in ``valid'' padding mode, i.e. boundaries are removed from the output.  This means that in each encoder and decoder stage the map loses some of its boundary information. To counter this effect, the input data (map and mask) needs to be extended periodically with just enough pixels so that the final output map has the correct size, here $128 \times 128$ pixels. This padding would not be needed if we had a convolution implementation on the sphere which would enforce periodicity automatically. In the decoder stages the map is upsampled by a factor of two before applying the convolution. Because of the boundary effects, the skip connection outputs shown in Fig.\ref{fig:fig_nn_1and2} need to be cropped on the boundaries before concatenating them to the decoder outputs. Note also that the full architecture has both a linear and a non-linear encoder-decoder path, while here we have only tabulated the linear path. The non linear path is equivalent except for the lack of a decoder 0. The number of filters of decoder 0 depends on whether we Wiener filter a T map (1 filter) or QU maps (2 filters).

In total, our architecture for temperature has about 80.000 free parameters, a small number compared to the neural networks usually deployed for image analysis, which have millions of parameters. We have found that for polarisation using a larger number of filters improves the results on large scales by a few percent. Our polarisation WienerNet has $620.000$ parameters, although the smaller temperature network also gives good results. In both cases the neural network provides an efficient representation of the Wiener filtering matrix, compared to a dense representation with $N_{\mathrm{pix}}^2\simeq (3 \times 10^7)$ elements.

\begin{table}[hbt!]
\begin{center}
\begin{tabular}{l|l|l|l|l|l|l}
                                              & \textbf{input size} & \textbf{output size} & \textbf{kernel size} & \textbf{filter number T / pol} & \textbf{stride} & \textbf{skip conn. crop} \\ \hline
Encoder 0    & $384 \times 384$          &  $380 \times 380$         & $5 \times 5$     & 10 / 16 & 1 &    \\
Encoder 1    & $380 \times 380$          &  $188 \times 188$         & $5 \times 5$     & 10 / 16 & 2  &    \\
Encoder 2    & $188 \times 188$          &  $92 \times 92$         & $5 \times 5$     & 10 / 32 & 2  &    \\
Encoder 3    & $92 \times 92$          &  $44 \times 44$         & $5 \times 5$     & 10 / 32 & 2  &    \\
Encoder 4    & $44 \times 44$          &  $20 \times 20$         & $5 \times 5$     & 10 / 32 & 2  &    \\
Encoder 5    & $20 \times 20$          &  $8 \times 8$         & $5 \times 5$     & 10 / 32 & 2  &    \\
Decoder 5   & $8 \times 8$          &  $12 \times 12$         & $5 \times 5$     & 10 / 32 & 2  &   \\   
Decoder 4   & $12 \times 12$          &  $20 \times 20$         & $5 \times 5$     & 10 / 32 & 2  &  $(4,4)$  \\   
Decoder 3   & $20 \times 20$          &  $36 \times 36$         & $5 \times 5$     & 10 / 32 & 2  &  $(12,12)$  \\   
Decoder 2   & $36 \times 36$          &  $68 \times 68$         & $5 \times 5$     & 10 / 32 & 2  &  $(28,28)$  \\   
Decoder 1   & $68 \times 68$          &  $132 \times 132$         & $5 \times 5$     & 10 / 16 & 2  &  $(60,60)$  \\   
Decoder 0   & $132 \times 132$          &  $128 \times 128$         & $5 \times 5$     & 1 / 2 & 1  &  $(124,124)$  \\     
\end{tabular}
\end{center}
\caption{Encoder and decoder stages for our default architecture with 6 encoder-decoder stages and convolutions of size $5 \times 5$.}
\label{tab:galaxy_surveys}
\end{table}

\end{document}